# Hypothesis of a Mundane Solution to the Pioneer Anomaly

Steven M Taylor

In recent years, phenomena involving the astronomical measurement of Doppler shift have produced unexpected results. Specifically, the so-called Pioneer anomaly, which is an apparent sunward acceleration of two spacecraft associated with a blue shift on top of an overall red shift of $(5.99 \pm 1.33) \times 10^{-9}$ Hz/s remains unexplained.[1] Also, Doppler shift measurements of distant galaxies have yielded results which indicate not a deceleration, but an acceleration of the rate of expansion. This so-called acceleration has so far defied a conventional explanation.

Incorporating the all-angle relativistic Doppler shift formulae in the interpretation of astronomical Doppler shift may be able to at least in part explain these unexpected measurements. While Astronomical Doppler shift calculations typically consider photons as radial/longitudinal emissions, which are shot back toward the observer, some mechanism may be causing a canting of the emission toward an angle of emission away from such an orientation. Such emissions would be of a higher frequency and would help explain a lower than expected red shift associated with both the Pioneer spacecraft and possibly distant galaxies.

Could such a canting be explained by the curving of the path of photons by gravity? When pitching a ball to someone at a constant velocity, one would have to increase the angle pitched to reach the same target if the distance between target and pitcher is increased. This could be analogous to the Pioneer spacecraft increasing their distance to Earth within the Sun's gravity well, as well as being analogous to galaxies increasing in relative distance with time. The possibility of the universe taking on a greater radius and increasingly less severe curvature with time could also contribute to the effect.

The perceived acceleration of the expansion rate of the universe is a complex problem. But a look at the Pioneer anomaly suggests the possibility of a mundane solution. Assuming a transmission frequency of two gigahertz, and a velocity of 132,000 kph demonstrates that a very small shift can account for the anomalous blue shift of approximately $5.99 \times 10^{-9}$ hz/s. The all-angle formulae for relativistic Doppler shift is $v' = v_0 \gamma (1 - \cos\theta \beta)$ where $\beta = v/c$.[2]
The derivative with respect to $\theta$ is $dv'/d\theta = v_0 \gamma \beta \sin\theta$.
Where $dv'/d\theta = 5.99 \times 10^{-9}$ hz/sec.
Therefore every second a shift of $5.99 \times 10^{-9}$ hz $= v_0 \gamma \beta \sin d\theta$;
and $\sin d\theta = 5.99 \times 10^{-9}$ hz $(v_0 \gamma \beta)^{-1}$
Therefore, $d\theta = \arcsin 5.99 \times 10^{-9}$ hz $(v_0 \gamma \beta)^{-1}$. Therefore, to get a $dv'$ of $5.99 \times 10^{-9}$ Hz/sec would require a shift in emission angle of only
$1.40 \times 10^{-12}$ degrees every second. That is a rate of approximately $22 \times 10^3$ years per degree. ***So it would not take much of a related shift to account for the Pioneer anomaly.***

[1] H. Dittus, et al. *A mission to explore the Pioneer anomaly*, (arXiv:gr-qc/0506139v1 30 Jun 2005)
[2] Steven M Taylor, *Finding the mass-energy counterpart to the transverse Doppler shift*, (arXiv:physics/0509261 30 Sep 2005)